\documentclass[fleqn,usenatbib]{mnras}

\usepackage[T1]{fontenc}
\usepackage{ae,aecompl}

\usepackage{epsf,palatino,url,graphicx,wrapfig,array,setspace,amsmath,amssymb,fancyhdr,multirow,lscape,appendix,rotating}
\usepackage{graphics,epsfig,txfonts}

\usepackage{longtable}
\usepackage{amssymb}
\usepackage{xcolor}
\usepackage{color}
\usepackage{lipsum}
\usepackage{textcomp}
\usepackage{threeparttable}
\usepackage{enumerate}
\usepackage{placeins}
\usepackage{float}
\usepackage[normalem]{ulem}
\usepackage{soul}
\usepackage{blindtext}

\newcommand{\xmm}{{\it XMM-Newton}\xspace}

\newcommand{\swift}{{\it Swift}\xspace}

\newcommand{\ergs}[1]{$\times 10^{#1}$ erg s$^{-1}$}

\newcommand{\ulx}{M51\,ULX-7\xspace}

\newcommand{\eqb}{\begin{eqnarray}}
\newcommand{\eqe}{\end{eqnarray}}

\title[Magnetic field of M51 ULX-7]{M51 ULX-7: super-orbital periodicity and constraints on the neutron star magnetic field}

\author[G. Vasilopoulos et al.]{
G.~Vasilopoulos$^1$\thanks{E-mail: georgios.vasilopoulos@yale.edu},
S. K. Lander$^{2}$,
F. Koliopanos$^{3,4}$,
C. D. Bailyn$^{1}$
\\
$^1$Department of Astronomy, Yale University, PO Box 208101, New Haven, CT 06520-8101, USA \\
$^2$Nicolaus Copernicus Astronomical Centre, Polish Academy of Sciences, Bartycka 18, PL-00-716 Warsaw, Poland\\
$^{3}$CNRS, IRAP, 9 Av. colonel Roche, BP 44346, F-31028 Toulouse cedex 4, France\\
$^{4}$Universit{\'e} de Toulouse; UPS-OMP; IRAP, Toulouse, France \\
}

\date{Accepted XXX. Received YYY; in original form ZZZ}

\pubyear{2019}

\begin{document}
\label{firstpage}
\pagerange{\pageref{firstpage}--\pageref{lastpage}}
\maketitle

\begin{abstract}
In the current work we explore the applicability of standard theoretical models of accretion to the observed properties of \ulx. The spin-up rate and observed X-ray luminosity are evidence of a neutron star with a surface magnetic field of $2-7\times10^{13}$ G, rotating near equilibrium.
Analysis of the X-ray light-curve of the system (\swift/XRT data) reveals the presence of a $\sim$39 d super-orbital period. 
We argue that the super-orbital periodicity is due to disc precession, and that material is accreted onto the neutron star at a constant rate throughout it. 
Moreover, by attributing this modulation to the free precession of the neutron star we estimate a surface magnetic field strength of $3-4\times10^{13}$ G. The agreement of these two independent estimates provide strong constraints on the surface polar magnetic field strength of the neutron star. 
\end{abstract}




\section{Introduction}

The discovery of pulsating ultra-luminous X-ray sources (PULXs) demonstrated that neutron stars (NSs) can sustain accretion at a super Eddington rate, assuming observed fluxes translate to isotropic luminosities \citep[][]{2014Natur.514..202B}. 
This discovery has fueled a search that led to the discovery and study of more PULXs in recent years
\citep[e.g.][]{2018MNRAS.476L..45C,2017Sci...355..817I,2014Natur.514..202B,2016ApJ...831L..14F,2019arXiv190604791R,2019MNRAS.tmpL.104S}.
Furthermore, based on spectral similarities between pulsating and non-pulsating ULXs, there is now compelling evidence in favor of highly magnetized NSs being the engines of ULXs  \citep{2017A&A...608A..47K,2018ApJ...856..128W}.

In order to explain the super-Eddington luminosities of PULXs, it has been speculated \citep[][]{2015MNRAS.454.2539M} that the NSs in PULXs must have higher magnetic fields strengths ($B>10^{13}$ G), than NSs in typical X-ray pulsars \citep{2014MNRAS.437.3664H,2016ApJ...829...30C}. 
However, the radiative mechanisms of the NS accretion column \citep{2007ApJ...654..435B} have not yet been fully studied at super-Eddington accretion rates, where several assumptions break due to the accretion column geometry \citep{2017ApJ...835..129W}.
Nevertheless, in some PULXs (see NGC300 ULX1) steady accretion at super Eddington rates and a NS with typical magnetic field (dipole term; $\sim10^{12}$ G) can coexist \citep{2018ApJ...857L...3W,2018A&A...620L..12V,2019MNRAS.488.5225V}.

A mechanism that could affect the long term behaviour of PULXs is the transition from accretor to propeller regime as the accretion disc inner radius changes size and becomes larger than the NS corotation radius \citep{1975A&A....39..185I}. This transition is often seen during the decay of outbursts from transient X-ray pulsars \citep{2016A&A...593A..16T} like Be X-ray binaries (BeXRBs).
Among PULXs the best candidate system for observing such transition is M82 X-2, a system that has shown a bimodal distribution in its X-ray flux \citep{2016MNRAS.457.1101T}, 
while the NS exhibits episodes of both spin-up (short term) and spin-down (secular evolution) that are consistent with propeller transition \citep{2019arXiv190506423B}. 

Another important characteristic of ULXs is the presence of ultra-fast outflows (UFO) \citep[e.g.][]{2016Natur.533...64P,2018MNRAS.479.3978K,2018MNRAS.473.5680K}.
These outflows originate from the disc as radiation pressure becomes important \citep{2007MNRAS.377.1187P}.
Moreover, outflows should be optically thick \citep{2009PASJ...61..213A}, and thus allow radiation to escape from a hollow funnel, resulting in mild to moderate beaming \citep{2017MNRAS.468L..59K,2019MNRAS.485.3588K}.  
The radiation originating from the central source can be scattered through the funnel walls, thus dramatically changing the temporal and spectral signatures of the signal seen by an observer that views the funnel under different orientations \citep{2009PASJ...61..213A,2015MNRAS.447.3243M}.
In the case of magnetized NSs as central engines, a large fraction of the pulsed emission can be scattered though the funnel walls, resulting in a decrease of the pulsed fraction \citep{2017MNRAS.468L..59K}. 
Interestingly, models that account for changes of the geometry of the system \citep[e.g. due to precession:][]{2018MNRAS.475..154M,2019MNRAS.489..282M} have been proposed to explain the super-orbital modulations of the X-ray light-curves of ULX systems \citep{2017MNRAS.466.2236D,2018arXiv181010518M}. 

However, the driving engine behind the disc and outflow precession is still unclear, and could be a combination of different physical processes.
There are numerous torques that act on the accretion disc and can cause its distortion (i.e.~warping) and change in orientation (tilt of disc) \citep{1998A&A...331L..37S,1999A&A...348..917S}. 
These are: the tidal force from the companion massive star, the interaction with the magnetosphere of the NS, or the irradiation of the disc \citep{1996MNRAS.281..357P,2010MNRAS.401.1275F,2001MNRAS.320..485O}.
One can argue that the presence of super-orbital periodicity in the X-ray light-curves of ULX is not necessarily evidence of outflows. After all many X-ray binaries, like Her X-1, show such behaviour on similar timescales.

The basic observable quantities of PULXs are the flux and super-orbital variability often seen in X-rays, the NS spin as well as the spin-up/down rate. These observables should provide the basis for testing accretion theory in extreme accretion rates, and compare with predictions of classical (i.e.~not super-Eddington) models.  
A major question is how does the accretion disc behave given the high accretion rates, and how do the disc properties alter the torque acting onto the NS and the variations of the observed X-ray flux. 
In a recent study, \citet{2019A&A...626A..18C} provide theoretical semi-analytical calculations for the inner disc radius and mass accretion onto the NS during super-Eddington accretion, by taking into account advection of heat and mass loss due to a wind originating from the disc \citep[see also,][]{2019MNRAS.484..687M}. 

To put things into perspective, we are now at a stage where we can test standard accretion theory in PULXs, and compare with numerical solutions that take into account more detailed physical properties. But most importantly, by obtaining deep X-ray observations, and by performing long X-ray monitoring campaigns we can investigate different aspects of ULXs and obtain self-consistent solutions for their properties.
We can then answer questions like: do PULXs host NSs with unusually high magnetic fields? are observed X-ray fluxes enhanced by beaming or are a product of super-Eddington mass-accretion rate? are outflows a necessary requirement for super-orbital modulation in the X-ray light-curves of PULXs?   

A particularly interesting system is \ulx, a known ULX system located in the Whirlpool galaxy at a distance of (8.58$\pm$0.1) Mpc \citep{2016ApJ...826...21M}. The detection of coherent pulsations ($P\sim2.8$ s) and an orbital period of $\sim2$ d confirm the system as a new PULX \citep{2019arXiv190604791R}. The X-ray spectral and temporal properties of the system have been presented by \citet{2019arXiv190604791R}.

In regard to the spin-evolution of \ulx, temporal analysis of \xmm data enabled the measurement of the NS spin in 4 epochs \citep{2019arXiv190604791R}.
According to the authors, between MJD 53552 and MJD 55723 the spin period of the NS evolved from 3.2831(2) s to 2.8014(7) s, yielding a secular spin-up of $\dot{P}=-2.57\times10^{-9}$ s/s. 
Between MJD 55723 and MJD 58284 the spin period of the NS evolved from 2.8014(7) to 2.7977148(2) s, yielding a secular spin-up of $\dot{P}=-1.67\times10^{-11}$ s/s. These intervals represent two epochs with different secular spin evolution, one being significantly slower than the other. 
The intrinsic spin-up of the system could only be constrained from the combination of two \xmm observations (span of 2 days), and was determined to be $\dot{P}=-2.4\times10^{-10}$ s/s (or $\dot{\nu}=3.1\times10^{-11}$ Hz/s) on MJD 58283.44.
Finally, based on the orbital solution obtained for the system, a lower limit on the mass of the companion has been placed at 8.3 $M_{\odot}$ \citep{2019arXiv190604791R}.
 
In regard to the spectral properties of \ulx,
\citet{2019arXiv190604791R} have performed spectral analysis on archival \xmm observations of the system. 
The system was detected during a hard high flux state and a soft low flux state.
For the observations obtained in the hard state, the spectra were similar and can be fitted by a dual thermal component \citep[see application to ULXs;][]{2017A&A...608A..47K,2018ApJ...856..128W}.
The soft component has a temperature of $\sim$0.4-0.5 keV and a size of 700-800 km
while the hard component has a temperature of 1.33-1.5 keV and a size of 90-100 km.
The fit to the X-ray spectra yields an unabsorbed X-ray luminosity ($L_{\rm X}$) of 5.6-7.1\ergs{39} (0.3-10.0 keV).  
For one observation (obs L; where L refer to low flux state) the spectrum was greatly different from the remaining observations and could be fitted by a single-temperature (keV$\sim$0.19 keV) black body component with a size of $\sim$1700 km.
The absorption corrected flux during this observation was estimated to be 3\ergs{38} (0.3-10.0 keV).
However, we note that \citet{2019arXiv190604791R} did not account for the possibility that the decrease in the observed flux is due to occultation. In general, it has been shown that the observed flux of ULXs can significantly underestimate the intrinsic X-ray luminosity if one does not account for the obscuration by outflows and/or a precessing disc \citep[e.g.][]{2019MNRAS.488.5225V,2018MNRAS.476L..45C,2015NatPh..11..551F}. 

In this work we will focus on the study of \ulx. In section \S \ref{sec:super}, we will present the X-ray light-curve of the system and discuss the presence of steady super-orbital periodicity. In section \S \ref{sec:accretion} we will discuss the basic predictions of standard accretion theory onto a magnetized NS, and use them to put constraints onto the magnetic field strength of the  NS in a self-consistent manner. Moreover, we will investigate whether our findings change by accounting for changes in the accretion disc structure due to the high accretion rate. We will do so by invoking the semi-analytical calculations of \citet{2019A&A...626A..18C}. We will show that \ulx cannot be in the so-called slow rotator regime, and the most probable solution is that the NS accretes at an almost constant accretion rate and is near spin equilibrium.
In section \ref{sec:prec} we will discuss the nature of the super orbital modulation, and argue that outflows are not a necessary condition for causing the observed pattern of variability, and a precessing disc could be a valid mechanism. A compelling engine that could drive the disc precession is free precession of the system's NS. This then leads to an estimate for how distorted the star is, and thus -- accounting for the superconducting state of the NS core -- enables us to estimate the NS magnetic field required to provide this distortion.

\section{Discovery of super-orbital periodicity}
\label{sec:super}

\begin{figure*}
  \resizebox{\hsize}{!}{
      \includegraphics[angle=0,clip=]{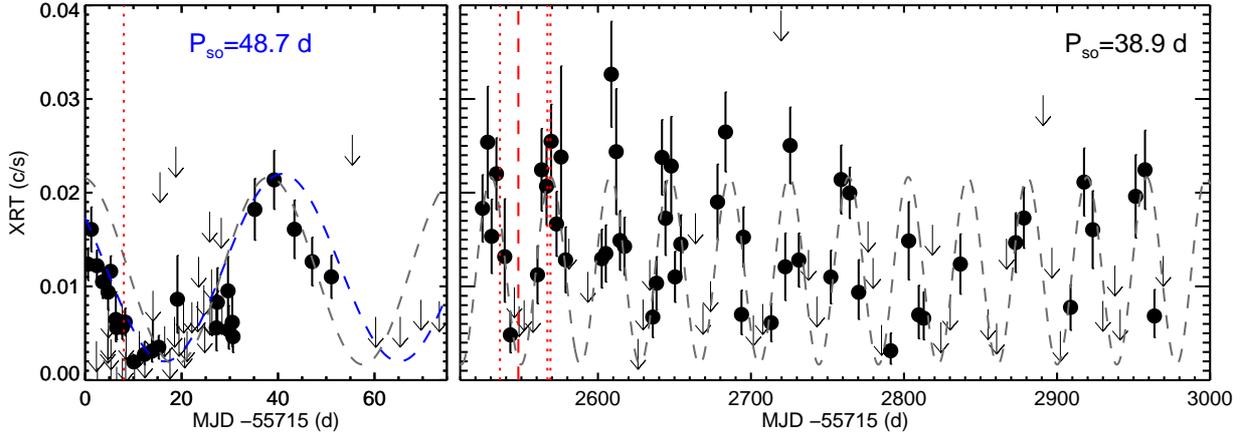}
     }
  \vspace{-0.55cm}
  \caption{X-ray light curve of \ulx based on \swift/XRT monitoring of the region for two different epochs. For non-detections, arrows represent 3$\sigma$ upper limits.
  Dashed gray sinusoidal curve is plotted to guide the eye, its 38.97 d period based on the periodicity detected during the second epoch (right plot). The curve is extrapolated to the earlier epoch for illustration purposes; we see that the early data points fit the curve well, hinting that the periodicity could have remained in phase for the intervening $\sim 60$ cycles for which we have no data.  However, given the putative periodicity  $\sim$48.7 d we find for the first epoch (see Fig. \ref{fig:LS}) and the inconsistency of the gray curve with the non-detections after MJD−55715 = 60 d, we also present a second curve (blue line) in better agreement with these findings. Vertical red lines mark the epoch where \xmm observations have been performed \citep{2019arXiv190604791R}. A super-soft spectrum has been detected during the observation marked with dashed red line (see text for details). For all \xmm observations marked with dotted red lines a NS spin period has been detected \citep{2019arXiv190604791R}.}
  \label{fig:xlc}
\end{figure*}

\begin{figure}
\vspace{-0.3cm}
  \resizebox{\hsize}{!}{
      \includegraphics[angle=0,clip=]{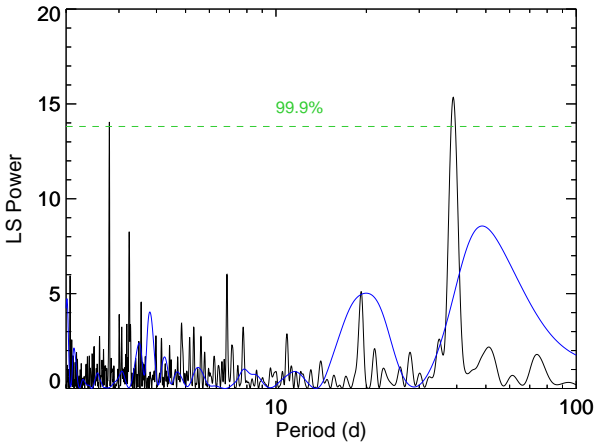}
     }
  \vspace{-0.5cm}
  \caption{Lomb-Scargle periodogram for epoch I (blue line) and epoch II (black line) as defined in Fig. \ref{fig:xlc}. Horizontal dotted line marks the 99.9$\%$ confidence level based analysis of simulated light-curves.}
  \label{fig:LS}
\end{figure}

\begin{figure}
\vspace{-0.3cm}
  \resizebox{\hsize}{!}{
      \includegraphics[angle=0,clip=]{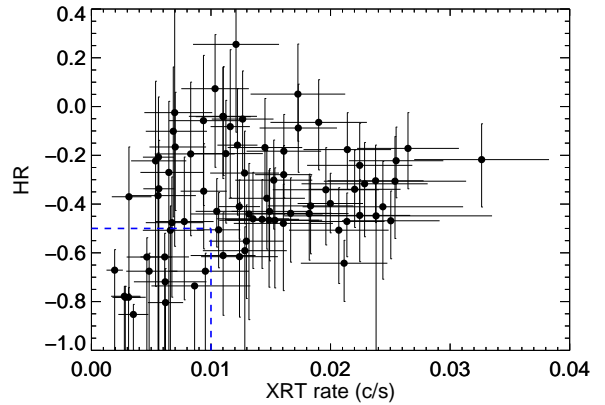}
     }
  \vspace{-0.5cm}
  \caption{Spectral evolution of \ulx as a function of the detected \swift/XRT count rate. Low count rates are consistent with a super-soft spectrum (i.e.~region marked with dashed blue lines).}
  \label{fig:HR}
\end{figure}

X-ray observations of M51 have been obtained by the
Neil Gehrels \swift\ Observatory \citep[\swift,][]{2004ApJ...611.1005G} X-ray Telescope \citep[XRT, ][]{2005SSRv..120..165B}.
To investigate the long-term variability of the system we searched for available \swift/XRT observations in the direction of M51. We found about 150 \swift/XRT observations, where $\sim$70 were performed between MJD 55715-55795 (Epoch I) and the remaining cover a span of about 500 days between MJD 58215-58715 (Epoch II).
We note that both epochs I and II correspond to the period of slow spin-up of \ulx as defined in the previous section (i.e.~MJD 55723-58282).
Data were retrieved and analyzed though the UK \swift \ science data centre\footnote{\url{http://www.swift.ac.uk/user_objects/}}, using standard procedures as outlined in \citet{2007A&A...469..379E,2009MNRAS.397.1177E}. 
To further investigate spectral changes we estimated the spectral Hardness Ratio (HR) from each observation.
HR is defined as the ratio of the difference over the sum of the number of counts in two subsequent energy bands: $\rm{HR}=(\rm{R}_{\rm{i+1}}-\rm{R}_{\rm{i}})/(\rm{R}_{\rm{i+1}}+\rm{R}_{\rm{i}})$,   
where $\rm{R}_{\rm{i}}$ is the background-subtracted count rate in a specific energy band. 
For this study we used the 0.3-2.0 keV and 2.0-10.0 keV bands.
Given the multiple X-ray sources in the \swift/XRT field we extracted source events from a 40\arcsec circular region, and background events from a combination of background regions from areas free from other sources. To account for the high background and the low number of source events during many observations we used a Bayesian estimator of HR \citep{2006ApJ...652..610P} that is publicly available as a command line program\footnote{\url{hea-www.harvard.edu/astrostat/BEHR/index.html}}. 

For the constructed \swift/XRT light curve we computed the Lomb-Scargle (LS) periodogram\footnote{Python code available at: \url{https://github.com/jakevdp/PracticalLombScargle/}} \citep{1982ApJ...263..835S,1986ApJ...302..757H,2018ApJS..236...16V}. When using the complete \swift/XRT light-curve we detected a periodic 
signal ($\sim$38.58d). A similar period ($\sim$38.87$\pm0.20$ d) with higher significance was derived when we used only the data from Epoch II. While the LS periodogram did dot yield any significant detection when using only data from Epoch I, the highest periodogram peak was found at $\sim$48.7 d. 
Given the \swift/XRT coverage within the first epoch, the 48.7 d period is quite uncertain, and we can only refer to this as an indication for a change in the super-orbital periodicity.

The X-ray light-curve of \ulx is presented in Fig. \ref{fig:xlc}. On the same figure we overplotted sinusoidal curves with a period measured by the LS periodogram. 
The LS periodogram for the 2 epochs is plotted in Fig. \ref{fig:LS}.
The computed HRs are plotted as a function of \swift/XRT count rate in Fig. \ref{fig:HR}.

Based on our analysis, we report on the presence of a quasi periodic signal in the X-ray light curve of \ulx. The period of modulation is $\sim39$ d, however there is some indication that the period can change and be as high as $\sim$49 d. During the low luminosity intervals the system seems to transit to a super-soft state where only emission from a soft thermal component is detected in deep \xmm observations. The estimated HRs support this claim as $\rm HR\xrightarrow{}-1$ for low count rates.

\section{Neutron star spin-up from accretion}
\label{sec:accretion}

\subsection{Basic equations}

Accretion theory has been the subject of numerous studies since the discovery of the first X-ray pulsars. For completeness and introducing the appropriate notations we will summarize the equations needed to constrain the magnetic field of the NS (i.e.~at polar surface $B$) from basic observable quantities. This will also help to remind the reader some of the basic assumptions in our treatment.

For an accreting object the Eddington luminosity is obtained by equating the outward radiation pressure with the gravitational force:
\begin{equation}
L_{\rm Edd}\approx1.5\times10^{38} m_1 \, {\rm erg \ s}^{-1},
\label{eq1}
\end{equation}
where $m_1\equiv M_{\rm  NS}/M_{\odot}$ is the NS mass (in units of the solar mass). 
At low mass accretion rates, the accretion disc around the compact object is locally sub-Eddington and its inner part is gas-pressure dominated \citep[e.g. see][]{2019A&A...626A.106M}. For mass accretion rates exceeding the critical rate $\dot{M}_{\rm Edd}$, part of the dissipated energy is used to launch mass outflows from the inner part of the accretion disc, thus leading to a reduced accretion rate onto the compact object. This critical luminosity is defined as:
\begin{equation}
\dot{M}_{\rm Edd}\approx 2\times 10^{18} m_1\, {\rm g \ s}^{-1},
\label{eq2}
\end{equation}

Outflows are launched inside the spherization radius $R_{\rm sph}$, where the disc thickness becomes comparable to its radius \citep{1973A&A....24..337S} or, equivalently, the disc  luminosity becomes equal to $L_{\rm Edd}$ \citep[see eq.~18 in][]{2007MNRAS.377.1187P}:
\begin{equation}
R_{\rm sph} \approx 10 \frac{GM_{\rm NS}\dot{m}_0}{c^2} \simeq 15 \, m_1\dot{m}_0 \, {\rm km},
\label{eq3}
\end{equation}
where $G$ is the gravitational constant, and $\dot{m}_0$ is the mass accretion rate at $R_{\rm sph}$ in units of $\dot{M}_{\rm Edd}$.
In a simple way, outflows are re-configuring accretion in a way that the local disc  accretion rate is sub-Eddington, and the mass accretion rate at $R<R_{\rm sph}$ can be  written as \citep{1973A&A....24..337S}:
\begin{equation}
\dot{M}(R)\simeq\frac{R}{R_{\rm sph}}\dot{m}_0\dot{M}_{\rm Edd}. 
\label{eq4}
\end{equation}
This approximation does not take into account the effects of heat advection in the disc. The latter results in a less abrupt decrease of $\dot{M}$ with radius than the one dictated by eq.~(\ref{eq4}) \citep[][]{2007MNRAS.377.1187P, 2019A&A...626A..18C,2019MNRAS.484..687M}. 

For magnetized NSs the accretion disc does not extend near the NS, but is truncated at much larger radii due to the interaction with the NS magnetosphere. The magnetospheric radius provides an estimate of the disc  inner radius \citep{1977ApJ...217..578G}:
\begin{equation}
R_{\rm M} = \xi \left(\frac{R_{\rm NS}^{12}B^4}{2GM_{\rm NS}\dot{M}^2}\right)^{1/7},
\label{eq5}
\end{equation}
where $R_{\rm NS}$ is the neutron star radius and $\xi\sim 0.5$ \citep{2018A&A...610A..46C}.
For typical $B$ values in X-ray pulsars (e.g., $10^{12}$ G), very high mass accretion rates are required 
(e.g., $\dot{m}_0>10$) to make $R_{\rm sph}>R_{\rm M}$. Moreover, as we will discuss in the following sections, the $\xi$ parameter could change as the mass accretion rate becomes super-Eddington, due to changes in the disc structure and conditions \citep[see for details:][]{2019A&A...626A..18C} .

An important condition in order for accretion to occur is that the Keplerian disc should be truncated inside the corotation radius:
\begin{equation}
R_{\rm co} = \left(\frac{G~M_{\rm NS}P_{\rm NS}^2}{4\pi^2}\right)^{1/3}.
\label{eq6}
\end{equation}
Let us now consider the interaction of the in-falling matter with the NS. As material is transferred from the inner disc radius to the NS, it also transfers angular momentum. 
The applied torque due to mass accretion is proportional to $\dot{M}_{\rm NS}$ (same as $\dot{M}(R_{\rm M})$) and $\sqrt{R_{\rm M}}$, as it acts like a lever arm onto the star (i.e.~$N_{\rm acc}\approx\dot{M}\sqrt{GMR_{\rm M}}$). 
An additional torque term acts due to coupling of magnetic field lines and the disc \citep{1979ApJ...234..296G}.
The total torque can be expressed in the form of $N_{\rm tot}=n(\omega_{fast})N_{\rm acc}$ where $n(\omega_{\rm fast})$ is a dimensionless function, and $\omega_{\rm fast}=(R_{\rm M}/R_{\rm co})^{3/2}$ is known as the fastness parameter.
\citet{1995ApJ...449L.153W} has provided an analytic relation for expressing this dimensionless function:
\begin{equation}
n(\omega_{\rm fast})=\frac{7/6-(4/3)\omega_{fast}+(1/9)\omega_{\rm fast}^2}{1-\omega_{\rm fast}}.
\label{eq7}
\end{equation}
Based on eq. \ref{eq7}, the induced torque onto the NS goes to zero as $\omega_{\rm fast}\rightarrow0.95$, and thus for $\omega_{\rm fast}>0.95$ the NS will spin down\footnote{We refer the reader to the recent work of \citet{2016ApJ...822...33P} for more details and a comparison between different regimes and solutions.}. 
For $R_{\rm M}{\simeq}R_{\rm co}$, the NS will spin down as the negative torque from interaction of the magnetic field lines with the accretion disc (that rotates slower than the NS) will dominate.  
For slow rotators (i.e.~$R_{\rm M}<<R_{\rm co}$) a valid approximation is $n(\omega_{\rm fast})=7/6$.
The change of the frequency of the NS is then given by: 
\begin{equation}
\dot{\nu}=\frac{n(\omega_\mathrm{fast})}{{\rm 2 \pi} I_{\rm NS}} \dot{M}(\rm R_{\rm M}) \sqrt{G M_{\rm NS} R_{\rm M}},
\label{eq8}
\end{equation}
where $I_{\rm NS} \simeq (1.0-1.7)\times10^{45}$~g cm$^{2}$ is the moment of inertia of the NS \citep[e.g.,][]{2015PhRvC..91a5804S}.
Equations \ref{eq6} and \ref{eq8}, assume angular momentum is transferred by a Keplerian disc. It has been speculated that the induced torque might be suppressed by radiation-pressure supported sub-Keplerian discs \citep{2005MNRAS.361.1153A}. However, detailed calculations have shown that non-Keplerianity should only affect the disc rotation by less than 10\% for typical parameters of PULXs \citep{2017MNRAS.470.2799C}.

The above equations are needed to probe the physical properties of a ULX system from basic observational quantities, like the spin-up rate and the observed $L_{\rm X}$. 
The measured $L_{\rm X,iso}$ of the system, can be converted to a mass accretion rate $\dot{M}$ assuming some efficiency.
This is assumed to be the efficiency with which gravitational energy is converted to radiation $L_{\rm X,iso}=GM_{\rm NS}\dot{M}/R$. For $R=R_{\rm NS}=10^6$~cm and $M_{\rm NS}=1.4M_{\odot}$, we find $L_{\rm X}\approx0.2\dot{M}c^2$. 
However, in ULX systems, a low to moderate amount of beaming can be present due to the presence of outflows \citep{2017MNRAS.468L..59K}. Thus the intrinsic luminosity ($L_{\rm X}$) of the system can be expressed as a function of beaming parameter $b$ (i.e.~$L_{\rm X}=L_{\rm X,iso}/b$):
\begin{equation}
\frac{L_{\rm X,iso}}{b}=\frac{GM_{\rm NS}\dot{M}}{R}
\label{eq9}
\end{equation}

To put the above into perspective, it is crucial to consider all above conditions and equations when probing the properties of \ulx. For example, if we invoke beaming in order to explain the observed super-Eddington luminosity, the derived mass accretion rate and $B$ should yield an inner disc radius smaller than $R_{\rm sph}$, as this is a necessary condition to have outflows and beaming \citep{2007MNRAS.377.1187P}.

\begin{table}
\begin{center}
\caption{Properties of \ulx}
\label{tab1}
\begin{threeparttable}
\begin{tabular*}{\columnwidth}{p{0.20\columnwidth}p{0.15\columnwidth}p{0.15\columnwidth}p{0.15\columnwidth}p{0.18\columnwidth}}
\hline\noalign{\smallskip}
parameter & \multicolumn{3}{c}{Values} & Units \\
\hline\noalign{\smallskip}
b factor$^{(a)}$  & 1 & 4 & 12 & \\
$\dot{M}_{\rm NS}$  &14.1 &  3.5  &  1.2 & $\dot{M}_{\rm Edd}$\\
$R_{\rm sph}$ $^{(b)}$  & 296 & 74 & 25 & km\\
\hline\noalign{\smallskip}
\multicolumn{5}{c}{Secular spin-up $^{(c)}$} \\
\hline\noalign{\smallskip}
 $B$ & $3\times10^{11}$& $1.9\times10^{12}$ & $5.2\times10^{14}$& G \\
 $R_{\rm M}$ & 140 & 2200 &  20000& km\\
\hline\noalign{\smallskip}
\multicolumn{5}{c}{Measured spin-up $^{(d)}$} \\
\hline\noalign{\smallskip}
 $B$ & $1.3\times10^{8}$  & $8.6\times10^{9}$ & $2.3\times10^{11}$ & G\\
 $R_{\rm M}$ & 1.7 & 28 & 248 & km\\
\hline\noalign{\smallskip}
\multicolumn{5}{c}{Comparison to other work $^{(e)}$ }\\
\hline\noalign{\smallskip}
 $B_{\rm AC}$ & $8.5\times10^{13}$& $1.2\times10^{13}$ & $8.5\times10^{11}$& G \\
 $R_{\rm M}$ &  3500 & 1700 &  520 & km\\
 $\dot{\nu}$ &  $1.4\times10^{-9}$ &  $2.5\times10^{-10}$ &   $4.6\times10^{-11}$ & Hz/s\\
\noalign{\smallskip}\hline\noalign{\smallskip}
\hline 
\end{tabular*}
{
 \tnote{(a)} Beaming factor used to translate the observed maximum $L_X=7.1\times10^{39}$ erg/s$^2$ to intrinsic luminosity and accretion rate. 
 \tnote{(b)} $R_{\rm sph}$ was estimated assuming no outflows; i.e $\dot{M}_{\rm NS}$ is the mass accretion rate of the disc. \tnote{(c)} The magnetic field was estimated based on the maximum observed secular spin-up rate; $\dot{\nu}=2.8\times10^{-10}$ Hz/s (see text). 
 \tnote{(d)} The magnetic field was estimated {\bf from eq. \ref{eq10},} based on the observed spin-up rate from individual observations $\dot{\nu}=3.1\times10^{-11}$ Hz/s (see text).
 \tnote{(e)} \citet{2019arXiv190604791R} have compared the maximum observed $L_X$ for a given beaming factor with the theoretical model of \citet{2015MNRAS.454.2539M} in order to derive the quoted $B$ values. 
}
\end{threeparttable}
\end{center}
\end{table}

\subsection{NS magnetic field from spin-up}

Having laid out the framework for estimating the NS magnetic field, we can proceed with an application to \ulx.
Analysis of archival \xmm data showed two periods with very different secular spin evolution \citep{2019arXiv190604791R}.
According to the authors, between MJD 53552 and MJD 55723 the spin period of the NS evolved from 3.2831(2) s to 2.8014(7) s, yielding a secular spin-up of $\dot{P}=-2.57\times10^{-9}$ s/s (or $\dot{\nu}=2.8\times10^{-10}$ Hz/s). 
Between MJD 55723 and MJD 58284 the spin period of the NS evolved from 2.8014(7) to 2.7977148(2) s, yielding a secular spin-up of $\dot{P}_{\rm sec}=-1.67\times10^{-11}$ s/s (or $\dot{\nu}_{\rm sec}==2.13\times10^{-12}$ Hz/s).
The spin-up of the system can also be constrained by using accelerated epoch folding. This was only feasible from the combination of two \xmm observations (span of 2 days), and was determined to be $\dot{P}=-2.4\times10^{-10}$ s/s (or $\dot{\nu}=3.1\times10^{-11}$ Hz/s) on MJD 58283.44.
The latter $\dot{P}$, is calculated by taking into account the orbital motion of the binary \citep{2019arXiv190604791R}. On the other hand, the secular spin-up is affected by Doppler shifts due to orbital motion. However, the effect of orbital modulation in the observed P value only has an amplitude of $\sim$0.006 s, thus the estimated $\dot{P}_{\rm sec}$ is only affected by $<$2\%.
The estimation of $B$ must be based on the simultaneous measurements of $L_{\rm X}$ and $\dot{\nu}$. Nevertheless, constraints can also be made by using the secular spin-up rate of \ulx, under certain assumptions. 

Assuming \ulx is at the slow rotator regime ($R_{\rm M}<<R_{\rm co}$) and the NS away from spin equilibrium, we can estimate $B$ at the NS surface and $\dot{M}$ onto the NS. 
For our calculations we adopt $\xi=0.5$, $R_{\rm NS}=10^6$ km, $M_{\rm NS}=1.4 M_{\odot}$, $I_{\rm NS}=1.3\times10^{45}\textrm{g cm}^{-2}$. 
By solving eq. \ref{eq5}, \ref{eq8}, \ref{eq9} for the polar magnetic field strength:
\begin{equation}
B \simeq 740 n(\omega_\mathrm{fast})^{-7/2} \xi^{-7/4} G^{3/2} M_{\rm NS}^{3/2} R_{\rm NS}^{-6} I_{\rm NS}^{7/2}  \dot{\nu}^{7/2}(L_{\rm X,iso}/b)^{-3} G,
\label{eq10}
\end{equation}
where $L_{\rm X,iso}$ is the observed luminosity, and $\dot{\nu}$ the observed spin-up rate. Then by using eq. \ref{eq2}, \ref{eq3}, \ref{eq5}, \ref{eq9}, we can estimate the $R_{\rm M}$ and $R_{\rm sph}$ for different beaming factors. Our numerical calculations are based on the \citet{1995ApJ...449L.153W} approximation (see eq. \ref{eq7}) and are presented in Table \ref{tab1}. 
Our results can be compared to those of \citet{2019arXiv190604791R}.
In their work the authors assumed that the maximum observed $L_{\rm X}$ from the system is related to the surface magnetic field strength of NS based on theoretical predictions; i.e.~$L_{\rm 39}\approx0.35B_{12}^{3/4}$ \citep{2015MNRAS.454.2539M}.
For convenience we refer to the theoretical predicted magnetic field of the accretion column as $B_{\rm AC}$.

From the values in Table \ref{tab1}, it is clear that no combination of parameters can self-consistently describe the observed values of \ulx. 
Specifically, by assuming no beaming, the estimated $B$ value is too small to explain the high luminosity from the accretion column \citep{2015MNRAS.454.2539M}. In addition, the intrinsic spin up rate yields inner disc radii consistent with the NS size, thus this solution is highly improbable. 
On the other hand, a moderate beaming (i.e.~4-12) yields a size for the inner disc much larger than the $R_{\rm sph}$, i.e.~is in contrast with the conditions needed for beaming. Finally, in some cases the estimated $R_{\rm M}$ is larger than the corotation radius of the system ($R_{\rm co}\sim3300$ km), thus indicating that our assumption for being at the slow rotator regime is not met. 
Therefore, at least some of our original assumptions behind this approach must be incorrect.
For example the structure of the disc in these high accretion rates should affect our estimations, or perhaps the system is not in the slow rotator regime.

\subsubsection{Comparison with semi-analytical models}

\begin{figure}
  \resizebox{\hsize}{!}{
      \includegraphics[angle=0,clip=]{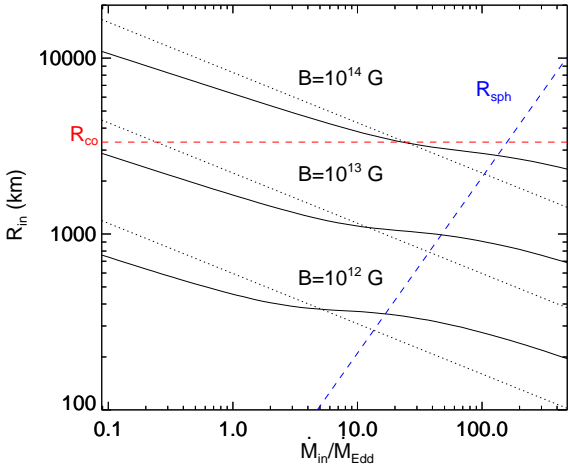}
     }
  \vspace{-0.5cm}
  \caption{Theoretical prediction of the dependence of the inner disc radius on the accretion rate at the inner disc \citep{2019A&A...626A..18C}. Standard solutions (i.e.~$R_{\rm in}\propto \dot{M}_{\rm in}^{-2/7}$) for $\xi=0.5$ are plotted with dotted lines. Blue dashed line follows  $R_{\rm sph}$, and horizontal red dashed line marks $R_{\rm co}$.}
  \label{fig:anna1}
\end{figure}

The analytical estimations of the previous section rely heavily on the assumption of a thin disc. Moreover, $\xi$ is known to vary from the default value of 0.5, for discs accreting at super-Eddington rates.
In these section we compare the observed properties of \ulx with the semi-analytical calculations of \citet{2019A&A...626A..18C}.  
In their recent work the authors have computed the inner radius of the accretion disc as a function of accretion rate for various $B$ values. These theoretical predictions are plotted in Fig. \ref{fig:anna1}. 
From Fig. \ref{fig:anna1} it is clear that for $B>10^{14} G$ a mass accretion rate larger than $10\dot{M}_{\rm Edd}$ is required for the system to be in the accretor regime. Moreover, for outflows and thus beaming to occur, the mass accretion at the inner disc radius (thus in the NS too: $\dot{M}_{\rm NS}$) should be larger than $10\dot{M}_{\rm Edd}$ even for a moderate magnetic field strength $10^{12} G$.

A comparison of the inner disc radius as estimated by the semi-analytical model of \citet{2019A&A...626A..18C} with the classical prescription provided by eq. \ref{eq5} can be used to provide an order of magnitude estimate for the uncertainties introduced by adopting the classical relation. For example there is always a specific $\dot{M}$ value where the two prescriptions provide the same result (see Fig. \ref{fig:anna1}). For smaller (larger) $\dot{M}$ values the semi-analytical estimations yield smaller (larger) $R_{\rm in}$ values for the disc. This difference can in general be no greater than a factor of 2 for typical $\dot{M}$ and $B$ values of PULXs.

\begin{figure}
  \resizebox{\hsize}{!}{
      \includegraphics[angle=0,clip=]{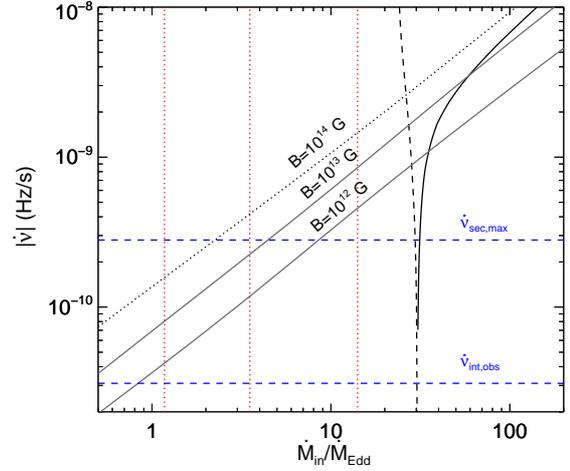}
     }
  \vspace{-0.5cm}
  \caption{Theoretical prediction of the induced spin-up of the NS as a  function of the accretion rate at the inner disc \citep{2019A&A...626A..18C}. Blue dashed line marks the observed maximum secular $\dot{\nu}$, and vertical red dotted line marks the calculated $\dot{M}_{\rm NS}$ for a beaming factor of 12, 4 and 1 (i.e.~no beaming) from left to right. In our calculation we have defined $n(\omega_{\rm fast})$  as in eq. \ref{eq7}. For magnetic field strengths of $10^{12}-10^{13}$ G, the system remains in the slow rotator regime for the range of accretion rates used in the plot. For $B=10^{14}$ G, spin equilibrium is reached for $\dot{M}_{\rm in}\sim30\dot{M}_{\rm Edd}$. For clarity we used solid and dotted lines to denote the spin-up and spin-down regimes respectively. Finally, with the dotted black line we plotted the expected spin-up rate if $\omega_{\rm fast}=0$ for $B=10^{14}$ G.}
  \label{fig:anna2}
\end{figure}

By using eqs. \ref{eq7} and \ref{eq8} we can compute the predicted NS frequency derivative and compare it with the observed values (see Fig. \ref{fig:anna2}). Given the spin of the NS ($\sim$2.8 s), the system should be in the slow rotator regime, for any $B<5\times10^{13} G$ and for a range of mass accretion rates larger than $0.1\dot{M}_{\rm Edd}$. However, for   
$B=10^{14} G$ the total torque should be zero for $\dot{M}_{\rm in}\sim30\dot{M}_{\rm Edd}$.

The large variations of observed secular $\dot{\nu}$ is also in favor of excluding solutions with low magnetic field (see Fig. \ref{fig:anna2}). For low $B$ the system should always remain in the slow accretor regime and variations of $\dot{\nu}$ by two orders of magnitude do not agree with mass transfer through Roche Lobe overflow \citep{2019arXiv190604791R}, that should be relatively stable. 
We conclude that the observed properties of \ulx cannot be adequately explained assuming a NS accreting in the slow accretor regime, even by using a more realistic disc structure compared to the standard thin disc.

Thus, following Occam's Razor the NS should be rotating near equilibrium, where relatively small changes in $\dot{M}_{\rm NS}$ should result in large changes in the induced torque.
For an equilibrium period of $P_{\rm NS}=2.8$ s and by following eq. \ref{eq7} we get a magnetic field of $4-7\times10^{13}$G assuming a beaming factor of $1-4$.

\subsection{Uncertainty of $B$ based on different torque models}

The estimated $B$ value is affected by our poor knowledge of accretion torque models. In the literature there are several torque models adopted for X-ray pulsars; i.e. \citet{1979ApJ...232..259G}, \citet{1995ApJ...449L.153W}, \citet{2004ApJ...606..436R}. 
Although most of them are in reasonable agreement for slow rotators, there can be a large discrepancy for the estimate of the equilibrium period. Analytically, this discrepancy lies in the derivation of $n(\omega_{\rm fast})$ that becomes zero for different $\omega_{\rm fast}$ values.
Following \citet{1995ApJ...449L.153W} the equilibrium period is found for $R_{\rm M}/R_{\rm co}\sim0.95$, while from the earlier prescription of \citet{1979ApJ...232..259G} we get $R_{\rm M}/R_{\rm co}\sim0.5$. \citet{2004ApJ...606..436R} focused in the case where $R_{\rm M}>R_{\rm co}$, and argued that in this case the disc re-configures such as $R_{\rm in}\sim R_{\rm co}$, while for cases where $R_{\rm in}<R_{\rm co}$ their solution is in reasonable agreement with the one by \citet{1995ApJ...449L.153W} \citep[see also][ and their Fig. 2 for a comparison]{2016ApJ...822...33P}. Based on \citet{1979ApJ...232..259G} prescription, and assuming 2.8 is the equilibrium period of \ulx, we get $B\sim2.3\times10^{13}$ G for the NS, or a factor of 3 lower value from the one derived using the \citet{1995ApJ...449L.153W} prescription.

Another mechanism that can potentially affect the spin equilibrium of the NS is the enhancement of spin-down torques from disc induced opening of the pulsar magnetic field lines \citep{2016ApJ...822...33P}. This mechanism has been used in order to naturally explain millisecond pulsars with spin periods larger than 1~ms, although classic torque theory would expect them to rotate at much higher rates. The opening of field lines could be responsible for enhancing the spin-down torque of the pulsar wind by a factor of $(R_{\rm LC}/R_{\rm M})^2$, where $R_{\rm LC}$ is the light-cylinder of the pulsar. 
The spin-down torque is thus given by \citep[see eq. 18 of][]{2016ApJ...822...33P}:
\begin{equation}
N_{\rm down,open}=\frac{(B^2R_{\rm NS}^3)^2}{R_{\rm M}^2}\frac{2\pi}{cP}, 
\label{eqP16}     
\end{equation}
where we have assumed maximum efficiency for the magnetic field opening.
Although this model is a decent approximation for the most rapid systems, for PULXs systems with high $\dot{M}$ the model predicts very high $B$ values, and yields equilibrium conditions that require $(R_{\rm co}<R_{\rm M})^2$. In physical terms, PULXs enter the propeller regime before this spin-down term becomes dominant. 
Thus, this mechanism alone could not be responsible for \ulx being at equilibrium.
For the case of \ulx, assuming all magnetic field lines between $R_{\rm LC}$ and $R_{\rm M}$ open, then the induced torques due to the enhanced pulsar wind would only equal to 20-25\% of the torque due to accretion \citep[see eq. 18 of][]{2016ApJ...822...33P}. 
As a result, it would effectively change the shape of the cusplike structure in the $\dot{M}$-$|\dot{\nu}|$ diagram (i.e. broaden it), and it would decrease the required magnetic field strength for making \ulx to be at equilibrium by $\lesssim3$\%.

To summarise, assuming no beaming in \ulx observed flux, uncertainties induced by our poor knowledge of torque models would yield a magnetic field strength of the NS of the order of $2-7\times10^{13}$G for an equilibrium period of 2.8 s.

\section{Nature of the super-orbital modulation}
\label{sec:prec}

The detection of a quasi-periodic super-orbital modulation in the X-ray flux of \ulx is perhaps a probe of precession from the disc.
Let us thus consider some basic mechanisms that could be responsible for such precession.

\subsection{Lense-Thirring precession}

\begin{figure}
  \resizebox{\hsize}{!}{
      \includegraphics[angle=0,clip=]{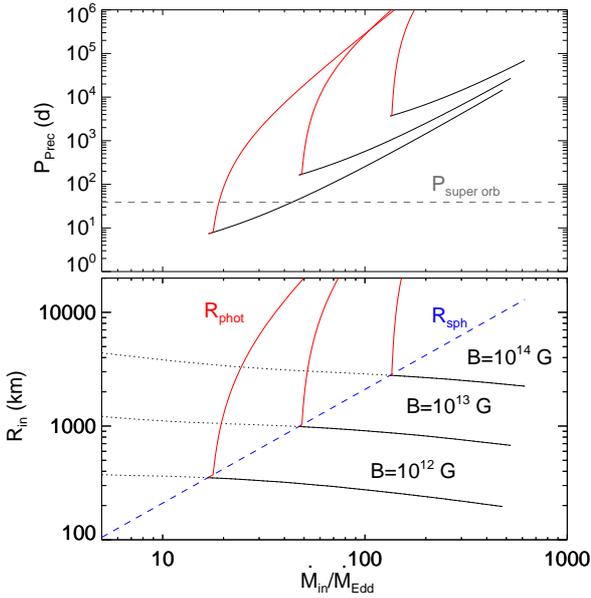}
     }
  \vspace{-0.5cm}
\caption{\emph{Lower panel:} Theoretical prediction of the inner disc radius as a function of accretion rate for various magnetic field strengths (same as Fig. \ref{fig:anna1}). A necessary condition for the presence of outflows is that the inner disc radius is smaller than $R_{\rm sph}$ (blue dashed line). The maximum extent of the outflows is given by the photospheric radius $R_{\rm phot}$ (red lines). \emph{Upper panel:} The precession period of the outflows can be estimated based on the model of \citet{2018MNRAS.475..154M}. A lower limit for the extent of the outflows, and thus for the precession period $P_{\rm Prec}$, can be given by $R_{\rm sph}$ (black lines); an upper limit for $P_{\rm Prec}$ (red lines) is obtained by assuming that the system precesses as a solid body up to a distance of $R_{\rm phot}$.
}
  \label{fig:prec}
\end{figure}

The effects of Lense-Thirring precession \citep{1975ApJ...195L..65B} have recently been discussed in the context of ULXs \citep{2018MNRAS.475..154M,2019MNRAS.489..282M}, in order to explain their super-orbital modulation. 
Moreover, it has been proposed to explain the rapid change of the radio jet orientation of V404 Cygni during its 2015 outburst \citep{2019Natur.569..374M}.
According to \citet{2018MNRAS.475..154M}, the precession period depends on parameters such as the NS spin, the size of $\rm R_{sph}$ and the extent of the outflows ($\rm R_{\rm out}$).
Following \citet{2018MNRAS.475..154M} the precession timescale is:
\begin{equation}
    P_{\rm prec}\approx P_{\rm NS}\frac{R^3_{\rm sph}c^2}{6 GI_{\rm NS}}\frac{1-(R_{\rm M}/R_{\rm sph})^3}{\ln{(R_{\rm sph}/R_{\rm M})}}\left(\frac{R_{\rm out}}{R_{\rm sph}}\right)^2,
\label{eq11}    
\end{equation}
An upper limit for the extent of the outflows can be estimated by estimating the distance at which the optical depth becomes less than one (i.e.~photospheric radius $\rm R_{\rm phot}$), while a lower limit is given by $\rm R_{\rm sph}$. 
The \citet{2018MNRAS.475..154M} model has been also used to explain the observed properties of NGC300\,ULX1 \citep{2019MNRAS.488.5225V}.
\citet{2019MNRAS.488.5225V} found that although the observed flux from NGC300\,ULX1 changed by a factor of $\sim50$ within a few months, the spin up of the system continued at a steady pace (constant $\dot{\nu}$). 
Based on the observed properties of NGC300\,ULX1 one can expect a timescale of precession of the order of a year, thus naturally explaining extended epochs of decreased flux due to obscuration from the outflows.

We can estimate the timescale of precession for \ulx by following \citet{2018MNRAS.475..154M} and invoking the same analytical relations used for NGC300\,ULX1 \citep[see eq. 13 \& 14 of ][]{2019MNRAS.488.5225V}. 
In Fig. \ref{fig:prec} we plot the predicted precession period as a function of mass accretion rate, for various values of $B$.
A timescale of precession similar to the observed one (i.e.~40 d), is only possible for super-Eddington accretion, and magnetic field strength lower than a few times $10^{12}$ G. However, this configuration requires \ulx to be in the slow-rotator regime and would yield a much higher NS spin-up rate than the observed one (see Fig. \ref{fig:anna2}).

\subsection{Precession due to the magnetic dipole torques}

The interaction of the disc and the magnetosphere is a complicated process that is not fully understood, but can lead to different quasi-periodic phenomena \citep{1999ApJ...524.1030L}.
The viewing angle of the disc can change on time-scales of months-years due to the interactions between the magnetic field and the disc \citep{1980SvAL....6...14L}. 
The spin-averaged torques by the dipole magnetic field lead to the inner twisted parts of the disc taking on a spiral-like shape with a certain direction of twisting. 
The forces applied to the disc will tend to twist it along the magnetic axis into a stable position.
The precession period is then given by:

\begin{eqnarray}
    P_{\rm prec,mag}&\approx& 1.5\times10^4\left(\frac{B}{10^{12} {\rm G}}\right)^{-2} \left(\frac{R_{\rm NS}}{10^6 {\rm cm}}\right)^{-2} \left(\frac{R_{\rm M}}{10^8 {\rm cm}}\right)^3 \left(\frac{P_{\rm NS}}{1 {\rm s}}\right)^{-1} \nonumber \\
    & & \left(\frac{I_{\rm NS}}{10^{45} \rm gr~cm^2}\right)\frac{1}{\cos{\psi}(3\cos{\zeta}-1)}~{\rm yr},
\label{eq12}    
\end{eqnarray}
where $\psi$ is the angle between the NS spin axis and the normal of the accretion disc, $\zeta$ is the angle between the spin axis and magnetic field axis.
Although this process has been proposed to explain the precession of PULXs \citep{2017MNRAS.467.1202M}, it would require extremely high (dipole) magnetic field values and mass accretion rates to reproduce the observed super orbital periods in PULXs (i.e.~$P_{\rm prec,mag}\propto{B^{-2/7}}\dot{M}^{-6/7}$).
For \ulx it would require mass accretion with a rate larger than $1000~\dot{M}_{\rm Edd}$ and $B$ higher than $10^{15}$ G. Thus a clock related to this mechanism would beat at much lower frequency than the observed super-orbital modulation. 

\subsection{NS free precession}
\label{sec:NSfree}

As we already mentioned, it is worth noting the similarities of \ulx with Her\,X-1, one of the first discovered X-ray pulsars. 
Her\,X-1 ($P=1.24$s) has an orbital period of 1.7 d, and a super-orbital period of $\sim$35 d \citep{1972ApJ...174L.143T,1973NPhS..246...87K}.
The nature of the super-orbital modulation has been speculated to to arise by precession of the companion star and a twisted accretion disc \citep{1975ApJ...201L..61P}, while a modulation of the pulse profile with super-orbital phase is also an indication of precession of the NS itself \citep{1986ApJ...300L..63T}.
NS free precession has also been invoked in order to explain long term periodic changes from observations of isolated NSs: both in their spectral properties \citep[e.g. RX\,J0720.4-3125][]{2006A&A...451L..17H} and in the long-timescale sinusoidal radio-timing residuals from a few sources \citep[e.g. PSR B1828-11][]{2000Natur.406..484S}.

Free precession does not necessarily act alone in any of these systems, however. For the case of Her\,X-1 it has since been argued that there are two different clocks that are responsible for its super orbital periodicities, and which are generally synchronized \citep{2009A&A...494.1025S}.
For the radio pulsar PSR B1828-11, the discovery of abrupt pulse-profile changes indicated the star's magnetosphere was switching between two different spin-down states and has led some to discard the precession model, since the apparently smooth slow oscillation of the residuals shows as discrete sharp steps on shorter timescales \citep{2019MNRAS.485.3230S}. On the other hand, the long-term average of these steps is so periodic that it seems that there must be some mechanism setting this periodicity -- and a freely-precessing NS could plausibly be such a clock \citep{2012MNRAS.420.2325J}.

In a similar way, the super-orbital periodicity of \ulx could be driven by free precession of the system's NS, even though we found indications that this period varies somewhat. Within this paradigm, assuming the possible accretion-disc precession to be synchronised with that of the NS, we can obtain useful constraints on any distortion $\epsilon$ of the star that is misaligned from the rotation axis (the motion is essentially independent of the axisymmetric centrifugal bulge of the star). This distortion may be either elastic or magnetic in origin.  Elastic distortions will be confined to the star's crust, although modelling of their size and structure is fraught with uncertainties, as they will depend on the star's seismic history; they could even be negligible. The star's internal magnetic field is a more reliable source of distortion, since a magnetic field always deforms its host star, with a predictable size scaling with the ratio of the magnetic energy to the gravitational binding energy \citep{1953ApJ...118..116C}.

Following \citep{2001MNRAS.324..811J} we get:
\begin{equation}
 \epsilon = \frac{P_{\rm NS}}{P_{\rm prec,obs}\cos{\chi}}\simeq7.72\times10^{-7} \frac{P_{\rm NS}}{1 \, {\rm s}} \left(\frac{P_{\rm prec,obs}}{30 {\rm d}}\frac{\cos \chi}{0.5}\right)^{-1} ,
\label{eq13}    
\end{equation}
where $\chi$ is the angle of the NS misaligned distortion compared to its rotation axis. For the observed parameters of the system we deduce $\epsilon=(0.83-1.2)\times 10^{-6}$ for the range\footnote{Note that equation \eqref{eq13} diverges in the limit $\chi\to 90^\circ$.} $\chi=0-45^\circ$.
 Now assume all this distortion is due to magnetic effects, so that $\chi$ represents the misalignment between the star's rotation and magnetic-field axes (and is therefore equal to the angle $\zeta$ from equation \eqref{eq12}). Quantitative solutions for a simple fluid model of a magnetically-distorted NS \citep{2009MNRAS.395.2162L} give:
\begin{equation}
 \epsilon = 2\times 10^{-11} \left(\frac{B_{\rm surf}}{10^{12}\textrm{ G}}\right)^2,
\label{epsB_normal}    
\end{equation}
where the $B^2$ scaling comes from the form of the magnetic energy. This leads us to infer a rather strong magnetic field, $B_{\rm surf}\approx 2\times 10^{14}$ G for the full range of $\epsilon$ we consider; \citet{2013MNRAS.435.1147P} obtained a similar estimate for Her X-1. However, this model is somewhat unrealistic, as the presence of superconducting protons in the NS core changes the magnetic-energy scaling to be proportional to $B H_{\rm c1}$, where $H_{\rm c1}$ is a `microscopic' magnetic field associated with the quantisation of magnetic flux on small scales. In practice, this leads to larger distortions for the same $B_{\rm surf}$, compared with those obtained from a non-superconducting model like equation \eqref{epsB_normal}. The ellipticity relation now takes the form  \citep{2013PhRvL.110g1101L,2014MNRAS.437..424L}:
\begin{equation}
 \epsilon = 3\times 10^{-8} \left(\frac{B_{\rm surf}}{10^{12}\textrm{ G}}\right)
                             \left(\frac{H_{\rm c1}(0)}{10^{16}\textrm{ G}}\right),
\label{epsB_sc}    
\end{equation}
where $H_{\rm c1}(0)$ is the value of the (density-dependent) $H_{\rm c1}$ at the centre of the star; we will set this to the sensible value of $10^{16}$ G. Using the above -- more physically reasonable -- ellipticity relation, our new estimate of the surface magnetic field is $B_{\rm surf}\approx (3-4)\times 10^{13}$ G for $\chi=0-45^\circ$.

Even if the full dynamics of the ULX system are complex, a magnetically-distorted, freely-precessing NS is an attractive candidate for the central clock governing its super-orbital modulation. 
For the mechanism to work, the NS must couple with the disc in an effective manner -- but the magnetospheric field lines provide a mechanism for this, tying the motion of charged particles in the disc with the NS on the Alfven crossing timescale of the system.
Note, however, that free precession of the NS alone cannot account for the putative variation in super-orbital period of $39-49$ days.

\section{Discussion}\label{sec:discussion}

Since the recent discovery of PULXs, 
the nature of their engines has been one of the major puzzles in high energy astrophysics. Highly magnetized NSs have been proposed as a mechanism for sustaining super-Eddington accretion \citep{2015MNRAS.454.2539M}. However, constant mass transfer from super-giant companions might also be a necessary requirement \citep{2019ApJ...878...71L}. Thus, observational constraints on the NS magnetic field and mass transfer are needed to understand these extraordinary systems.

The recent detection of super-orbital flux modulation in PULXs (M82~X-2: ${\sim}$60\,d, \citealt{2016MNRAS.461.4395K}; NGC 5907: ${\sim}$72\,d, \citealt{2016ApJ...827L..13W}; NGC~7793~P13: ${\sim}$65\,d, \citealt{2017ApJ...835L...9H}) has sparked renewed interest in this long-term periodic behavior, which is increasingly regarded as an indication for the presence of beamed emission in ULXs --  the result of a funnel-like structure, generated by massive outflows. However, it is important to recall that long-term (10-100\,d) periodic flux modulations (by a factor of ${>}$10), have long been detected in several normal X-ray pulsars with sub-Eddington luminosities (e.g., Cyg~X-2: \citealt{2000ApJ...528..410P}; Her~X-1: \citealt{1973NPhS..246...87K}; LMC~X-4: \citealt{2003A&A...401..265N}; SMC~X-1: \citealt{1981A&A....97..134B}).
Therefore, it could be argued that this type of variability in X-ray binary systems is simply an indication for the presence of a high-B NS, rather than super-Eddington accretion. 

Analysis of the \swift/XRT monitoring data of \ulx revealed the presence of a $\sim$39 d super-orbital modulation. 
Our calculations suggests that the coupling of the NS free precession and the disk through the magnetospheric torques constitutes a plausible mechanism for this behavior, as has been demonstrated for other X-ray pulsars with similar characteristics \citep{2013MNRAS.435.1147P}. We improve on previous modelling by accounting for the effect of superconductivity in the NS core on the precession period.
An attractive feature of the free-precession model is that its inherent uncertainties (like the value of $\cos\chi$ and the exact model leading to equation \eqref{epsB_sc}) only introduce order-unity changes, and -- unlike Lense-Thirring precession -- this model is independent of the complex magnetospheric physics of the system.

From the observational point of view, both Lense-Thirring and NS free precession timescales scale linearly with the NS spin. 
Thus, continuing monitoring of PULX systems like \ulx could provide strong arguments in favor or against each mechanism. 
Given that Lense-Thirring precession also strongly depends on $\dot{M}$, the NS free precession scenario can be tested in a more straightforward way.
The fact that most PULXs show changes in P on timescales of decades complicates such endeavor. Apart form \ulx, another potential candidate for such study might be the PULX NGC 300 ULX1, since in this particular system the NS is still at the slow rotator regime and its spin period has dramatically changed the last 4 years \citep[from $\sim$124 s to $\sim$16 s; see ][]{2018A&A...620L..12V,2019MNRAS.488.5225V}. However, in NGC 300 ULX1 no super-orbital period has been measured and given the system current properties (P$\sim$16 s) any periodicity due to precession might have timescales of years.

We must also note that several authors have proposed the precession of a radiation-driven warped disk as the origin of super-orbital variability in X-ray binaries \citep[e.g.][]{1996MNRAS.281..357P,1999MNRAS.308..207W,2001MNRAS.320..485O,2003MNRAS.339..447C,2019MNRAS.482..337D}. This scenario -- which has also been proposed to interpret similar temporal characteristics in ULXs M82~X-2 \citep{2016MNRAS.461.4395K} and SS~433 \citep{2010MNRAS.401.1275F} -- cannot be ruled out for \ulx. 
Nevertheless, further investigation of the precessing warped disk mechanism as the engine behind the super-orbital modulation can be performed by smoothed particle hydrodynamic simulations and not by analytical or empirical relations. This is beyond the scope of this work and will be the focus of a future investigation.

The comparison of the observed properties of \ulx with theoretical models has demonstrated that strong beaming ($b>2$) can be excluded, since a necessary condition should be the presence of outflows (i.e.~$R_{\rm M}>R_{\rm sph}$; see Fig. \ref{fig:anna1}). Moreover, although the NS exhibits a relative high secular spin-up rate, $|\dot{\nu}|$ is much smaller than would be expected for a system in the slow rotator regime (see Fig. \ref{fig:anna2}). 
One obvious explanation is that the NS is rotating near equilibrium, in which case its magnetic field should be $2-7\times10^{13}$ G.
This is in good agreement with the independent estimate of $B$ from NS free precession (see section \ref{sec:NSfree}).
In the free-precession paradigm, the torque acting by the magnetosphere on the disc will change in phase with the NS precession. Although the coupling of the disc and NS precession period is a complicated process, free precession could be the clock behind the observed super-orbital modulation \citep{1999ApJ...524.1030L,2013MNRAS.435.1147P}.
In the case of an accretion disc precessing due to synchronization with the NS free precession, no strong outflows are required to explain the modulation in the X-ray light-curve.  

An alternative scenario, to account for the low spin-up rate,  would be that the NS rotation axis and the accretion disc plane are near-aligned, so that there is minimum angular momentum transfer from the disc to the NS.    
As the energy of the interaction of the disc and the magnetic dipole depends on their relative orientation, a perpendicular or parallel configuration between the rotation axis and the axis of the disc would provide a stable solution \citep{1980SvAL....6...14L}. 
However, an application and confirmation of such models requires constraints on more modeled parameters -- as the interactions between magnetized NSs and accretion discs are much more complicated-- and a complete reference to different models would be beyond the scope of this work \citep[see][ and references within]{2014EPJWC..6401001L,1999ApJ...524.1030L}.

Attributing the $39$ d modulation to disc precession can naturally explain the spectral changes observed with \xmm \citep{2019arXiv190604791R} and \swift/XRT monitoring (see Fig. \ref{fig:HR}). When the accretion disc is edge-on (low state) we only see the super-soft component from the disc and disc wind that create a photosphere \citep{2019ApJ...871..115Z}.
Thus, during its low state \ulx is an analog to the ultraluminous super-soft X-ray sources like NGC\,55\,ULX \citep{2017MNRAS.468.2865P}. 
Moreover, by invoking the NS free precession and a warped disc to explain the super orbital modulation in \ulx, it is not required the disc to be in the advection dominated regime, where outflows are expected \citep{2019A&A...626A..18C,2019A&A...626A.106M,2019MNRAS.484..687M}. Thus, $R_{\rm M}>R_{\rm sph}$ is not a necessary condition for this phenomenon to take place. 
We note that even for the case where the $R_{\rm M}>R_{\rm sph}$ a warping of the outer disc due to irradiation from the AC is quite likely to happen \citep{1996MNRAS.281..357P,2018MNRAS.475..154M}. However, the spectral properties during the low flux state (i.e. size and temperature of emitting region) are more compatible with emission from the innermost disk part.
The soft emission originates from reprocessing of the beamed pulsar emission by a warped inner accretion disc that precesses with time (SMC\,X-1, \citealt{2005ApJ...633.1064H}; LMC\,X-4, \citealt{2010ApJ...720.1202H}; SMC\,X-3, \citealt{2018A&A...614A..23K}). In this picture the pulsar is covered by a shell of material (i.e. warped disk), while the covering fraction is the luminosity ratio of the soft to hard states $\Omega\sim L_{\rm soft}/L_{\rm hard}\sim0.05$ \citep{2004ApJ...614..881H}. 
An order of magnitude estimate of the size of the reprocessing region can be obtained by the temperature of the soft component observed in the low state (i.e. 0.19 keV) and the isotropic $L_{\rm X}$ of the pulsar as derived from the high state (i.e. $\sim$7\ergs{39}) by following \citep{2004ApJ...614..881H}:
\begin{equation}
R_{\rm rep}^2 = \frac{L_{\rm X}}{4\pi\sigma T_{\rm soft}^4},     
\end{equation}
where $\sigma$ is the Stefan-Boltzmann constant. We then get $R_{\rm rep}\sim6000$~km, or about 2 times the corotation radius. This is more evidence in favor of the disc being truncated near corotation radius and thus the NS rotating near equilibrium period.

Finally, given the evidence of change in the period of the super-orbital modulation, and thus the precession period of the disc, it is crucial to obtain future monitoring observations of \ulx. If the precession period is well maintained it would be evident of a stable clock, and the low luminosity states would only vary as a random walk \citep{1983A&A...117..215S}.
However, a long monitoring could reveal drift in the start epoch of the low states similar to systems like Her\,X-1.

\section{conclusion}

We have investigated the applicability of standard theoretical models to the observed properties of \ulx. The spin-up rate and mass accretion rate derived from observations are evident of a NS with a magnetic field of $2-7\times10^{13}$ G rotating near equilibrium. Analysis of \swift/XRT archival data have shown the existence of a super-orbital period of $\sim$39 d. 
Changes in the X-ray flux and spectral shape within the super-orbital period are most probably associated with disc precession that causes occultation and partial obscuration of the NS. 
Thus, no change in the accretion rate onto the NS is required to account for the super-orbital variability and mass transfer between the binary seems to remain fairly constant in timescales of years.
If the super-orbital modulation is related to the NS free precession then the surface magnetic field strength of the NS should be $\sim3-4\times10^{13}$ G. The agreement of these two independent estimates provides a strong argument for the NS free precession being responsible for the precession of the accretion disc in ULXs hosting magnetized NSs. 

\section*{Acknowledgements}
The authors would like to thank the anonymous referee for the the constructive report that helped to improve the manuscript.
GV would like to thank M. Petropoulou for providing useful comments on the draft.
We acknowledge the use of public data from the \swift\ data archive.




\bibliographystyle{mnras}
\bibliography{general}








\bsp	
\label{lastpage}
\end{document}